\documentstyle[11pt,epsf]{article}

\begin{document}

\title{
{\bf QED(1+1) at Finite Temperature --  \\
 a Study with Light-Cone Quantisation.  }
}
\author{
Stephan Elser \\ 
  {\it Institut f\"ur Physik, Humboldt-Universit\"at} \\  
  {\it Invalidenstrasse 110, D-10099 Berlin, Germany }    
 \\  \\ 
Alex C. Kalloniatis\thanks{Present Address: Institut f\"ur
Theoretische Physik III, Universit\"at Erlangen-N\"urnberg,
Staudtstrasse 7, D-91058 Erlangen, Germany.} \\
  {\it Max-Planck-Institut f\"ur Kernphysik }\\  
  {\it Postfach 10 39 80,  
  D-69029 Heidelberg, Germany }   }
\date{10 January, 1996} 
\maketitle

\begin{abstract}
We explore quantum electrodynamics in (1+1) dimensions at finite
temperature using
the method of Discretized Light-Cone Quantisation.
The partition function, energy and specific heat are computed 
in the canonical ensemble 
using the spectrum of invariant masses computed with
a standard DLCQ numerical routine. 
In particular, the specific heat exhibits a peak which grows
as the continuum limit is numerically approached. 
A critical exponent is tentatively extracted. 
The surprising result
is that the density of states contains significant finite
size artifacts even for a relatively high harmonic resolution.
These and the other outstanding problems in the
present calculation are discussed. 
\end{abstract}

\vspace{2cm}
{\bf Preprint MPI-H-V39-1995}
\newpage
\section {Introduction}
Developing nonperturbative methods of performing
computations at finite temperature for hadronic
systems is soon to become in demand with the advent
of experiments such as the RHIC to explore the phase transition
to and properties of the putative Quark-Gluon Plasma.
For zero temperature, there is now much activity in
also developing Dirac's front form \cite{Dir49} or Hamiltonian light-cone
field theory for studying nonperturbative field theories
\cite{PaB85} and extracting,
for example, the relativistic bound state spectra. 
Reviews can be found in \cite{BrP92}. Since the purpose
of this method is to write down a Hamiltonian, diagonalise it
and thus obtain a spectrum there would appear to be nothing
conceptual in stepping from there to the computation of
thermodynamic quantities: the spectrum, as the basis for 
developing the partition function, has -- at least in
principle -- all
the information one could require.

For this reason we tackle in the present work the finite
temperature properties of quantum electrodynamics in (1+1)
dimensions. Over several years and in a number of separate works
\cite{EPB87,Els94,EPK95}, 
the Hamiltonian and spectrum of the theory have been studied
with Discretized Light-Cone Quantisation (DLCQ). 
Comparison with what lattice data 
is available \cite{CrH80} has been
encouraging. It is also known that at least {\it some} 
-- though not all -- of the
physics of the fermion condensate in the small fermion
mass limit \cite{BSK76} are contained in the spectrum even with the trivial
vacuum \cite{Ber77,Els94}. QED(1+1) is thus the ideal place to 
test the light-cone
Hamiltonian method for finite temperature and even to make some
contribution to what little is known about the `massive Schwinger
model' at finite temperature. 

The main subtlety in the DLCQ program
is the continuum limit. One works at any stage in the numerical
work with a 
finite length interval manifested in momentum space
by a finite total momentum $K$, or {\it harmonic
resolution}, available for distribution
amongst the `partons' of the theory. Numerically, one
computes at various $K$-values
and, when feasible, extrapolates to extract the
continuum limit. For any finite $K$ the resulting spectrum is
finite and this of course has an impact on thermodynamic
quantities in the sum over all
eigenstates of the Hamiltonian. 
We study how this works its way into the various thermodynamic
quantities such as partition function, energy and specific heat. 
The partition function for various $K$-values varies enormously --
however this is not directly a physical quantity. We observe that in
the energy for low T the results are roughly $K$-independent
while above a certain temperature regime the energy becomes
{\it linear} in the temperature with the {\it slope}
being independent of $K$. This leads, in the specific heat,
to a $K$-independent result except in the neighborhood of 
$T= 0.5 - 1.0$, in units of $g/\sqrt{\pi}$,
where a peak with increasing height for increasing
$K$ appears. 
The growing
peak is suggestive of a property of the system as we approach
an infinite number of degrees of freedom.
We tentatively accept this to indicate a phase transition,
though it is not associated with the chiral condensate
as order parameter (which seems to be nonzero for
all finite temperatures anyway \cite{Smi92}). Rather, 
it appears to be related to the change in the 
spectrum from a discrete one of bound
mesonic states to a continuous spectrum of scattering states.

First, we briefly review the light-cone
formalism as applied to QED(1+1),  
discuss the basic spectrum and give the corresponding density
of states as determined in DLCQ. We work in the canonical ensemble
and present results for the partition function, energy
and specific heat for various values of harmonic resolution.    
The letter concludes with
a discussion of the results and the outstanding problems. 

\section{DLCQ for QED(1+1)}
The literature is already very extensive in references to the
basic method of DLCQ \cite{PaB85,BrP92}. We can therefore afford 
to be brief. Our light-cone
conventions are those of \cite{KoS70}.
Light-cone quantization involves initialising independent
quantum fields at equal light-cone time, say $x^+ = (x^0 + x^3)/\sqrt{2}=0$. 
The orthogonal $x^- = (x^0-x^3)/ \sqrt{2}$ is the longitudinal or light-cone
space. The appropriate time evolution generator is
$P^-$, itself a constant of the motion under evolution in $x^+$,
together with the momentum or $x^-$-translation generator $P^+$.  
Thus these operators can be built via the energy-momentum tensor
from the bare Lagrangian fields once
all redundant variables have been eliminated.
In order to regularise the infrared, a finite interval is
employed: $x^- \in \left[ -L ,+L \right]$. 
Bosons are assigned periodic and fermions antiperiodic boundary conditions. 
In the following we ignore zero modes \footnote{Their
complete incorporation into the analogue of these calculations is
still under investigation.}. Then the light-cone gauge $A^+=0$
is permissible in which the remaining $A^-$ gauge potentials  
are constrained and can be eliminated by solving the Gauss law
constraints. Thus, apart from the resulting linear Coulomb potential,
there is no trace left of the photons in two dimensional QED.
Similarly, only half the fermion degrees of freedom are independent,
the rest being eliminated by the Dirac equation -- a
strictly light-cone peculiarity. One is thus able
to completely specify the Poincar{\'e} generators in terms of the
so-called `right-mover' fermion field which is quantised canonically.
The vacuum is just the perturbative Fock vacuum in this framework
\cite{Wei66}. Hence states can be built up by application of
creation operators in the right-movers after an expansion in plane
wave modes. 
 
The relativistic bound-state problem
can be formulated as the   
eigenvalue equation 
\begin{equation}
2 P^+ P^- | \Psi_i \rangle = m^2_i |\Psi_i\rangle  
\;.
\end{equation}
Here, $P^\pm$ is expressed in terms of the Fock operators
into which the fields are expanded, and $|\Psi_i\rangle$ is
similarly built from Fock operators on the trivial vacuum. 
Inserting a complete set of Fock states  
results in a finite matrix equation
\begin{eqnarray}
\big<i\big| : 2 P^+ P^- : \big|j\big> \langle j \big|\Psi\big>
= m^2 \langle i \big|\Psi\big> ,
\end{eqnarray}
which has been diagonalized numerically \cite{EPB87,Els94}.
The reader can find the explicit form of the Hamiltonian
in \cite{EPB87,Els94}. We give here just the conventions
for scaling of the coupling and invariant mass.
With $g$ the gauge coupling and $m_F$ the
fermion mass in the QED(1+1) Lagrangian, we define
a coupling 
\begin{equation}
\lambda \equiv \left[ 1 + { {\pi  m_F^2} \over {g^2} } \right]^{-1}
\end{equation}
and a mass scale 
\begin{equation}
m_E^2 \equiv m_F^2 + { {g^2} \over \pi }
\;.
\end{equation}
The coupling $\lambda$ allows us to explore the weak
($\lambda \rightarrow 0$) to strong ($\lambda \rightarrow 1$)
coupling regimes within a finite range plot. 
In the following, all energies shall
be in units of $m_E$ as shall be the units of temperature
as well.  

On the computer it is more convenient to express the continuum
limit in terms of the `harmonic resolution' $K = L P^+/ \pi$ which is
related to the total (integer) momentum available to be
distributed to the partons of the theory. Evidently, in order to
take $L\rightarrow\infty$ with fixed total momentum one must
then take $K\rightarrow\infty$. For some physical quantities
it is feasible to compute for a range of values of $K$ and
extrapolate, however in the present work we are restricted
to studying thermodynamic quantities for finite but increasing
$K$. 
Herein lies one problem to which we
will have cause to refer frequently: computing with
$K\neq\infty$ generates an error
in the numerical results, however as yet this error lacks
a physical interpretation.
Nonetheless, 
the maximum value of $K$ chosen enables reproduction of results
for the low energy mass spectrum consistent with lattice gauge theory
but still permits computation on a workstation. The programs are
described in some detail in \cite{EPK95}.

\section{Spectrum and Density of States}
Before diving into the thermodynamics, we prepare ourselves
by reviewing the basic properties of the spectrum of QED(1+1).
A typical spectrum can be found in \cite{EPB87}. For completeness
we reproduce one used to calculate the quantities
of this letter; applying
slightly different boundary conditions as discussed in \cite{Els94}.

\begin{figure}[tbp]
\begin{center}

\begin{picture}(170,80)
\put(7  ,0)   {\bf a)}
\put(80 ,0)   {\bf b)}

\put(0,0){\unitlength1mm
 \begin{picture}(75,75)
  \centering
  \makebox[75mm]{
   \epsfxsize=75mm
    \epsffile{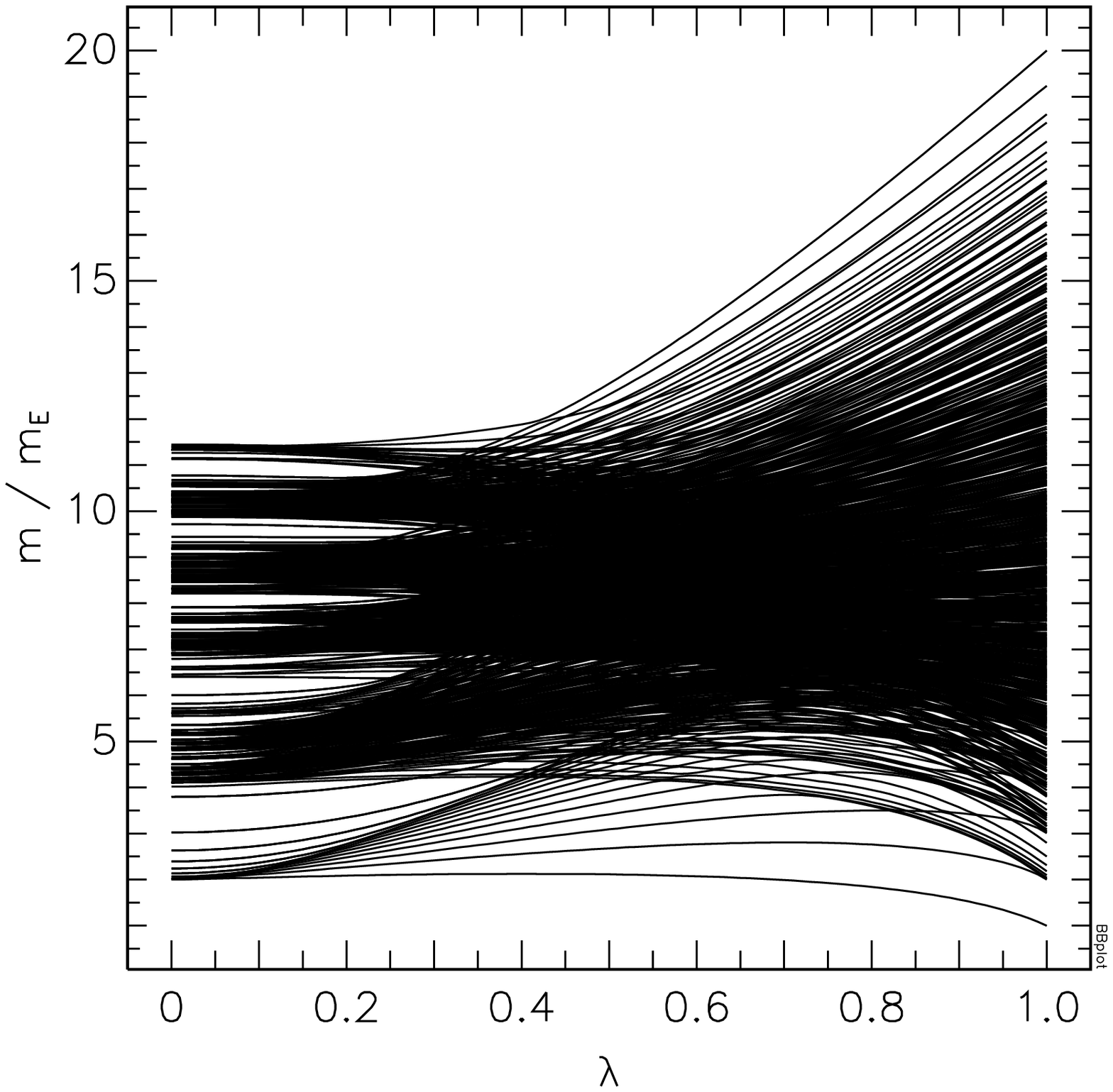}
  }
 \end{picture}
}
\put(75,0){\unitlength1mm
 \begin{picture}(75,75)
  \centering
  \makebox[75mm]{
   \epsfxsize=73mm
    \epsffile{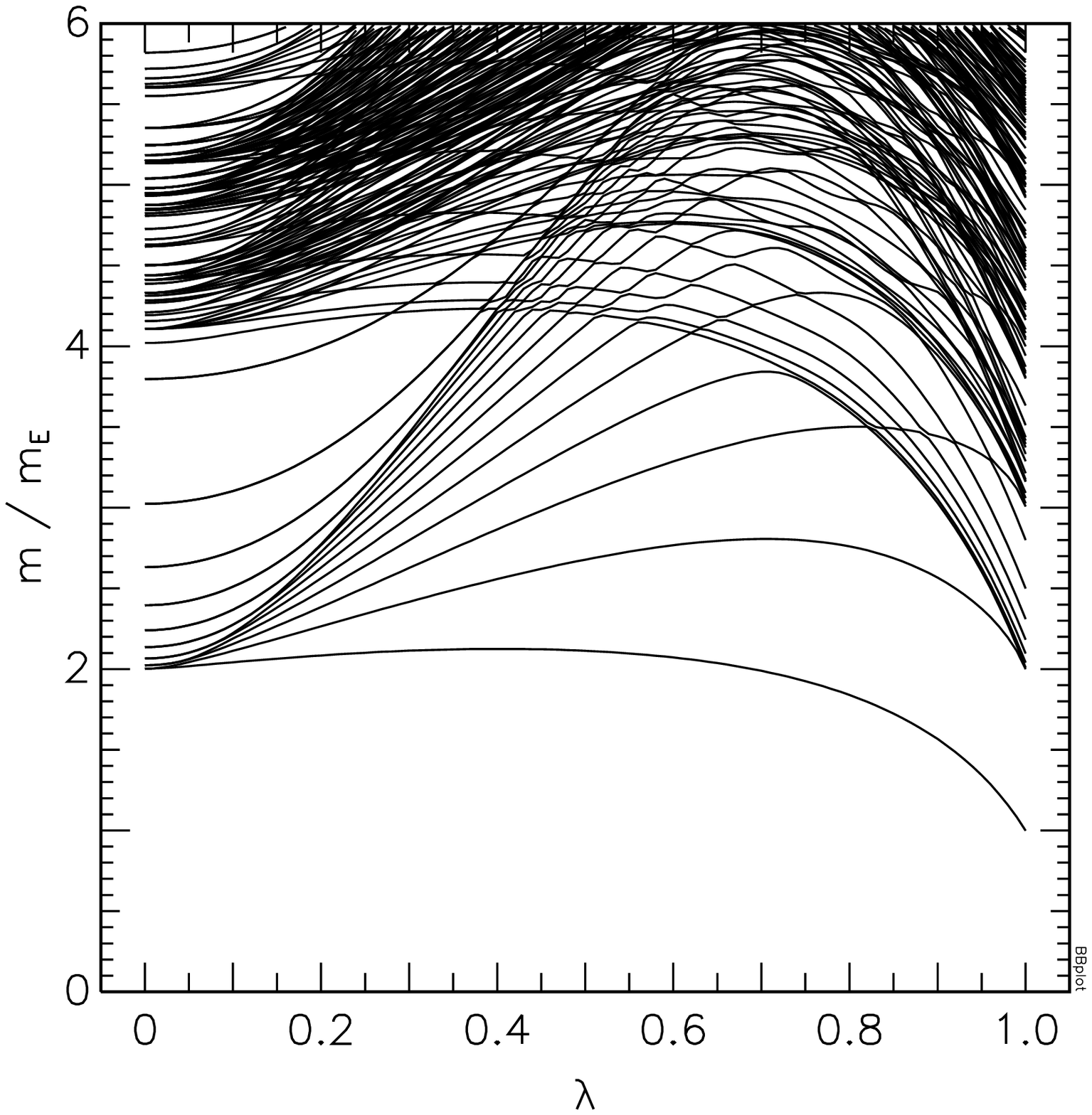}
  }
 \end{picture}
}

\end{picture}
\end{center}

\caption[Mass spectrum]
{{\bf Mass Spectrum from DLCQ.}
\sl
{\bf a)} The Mass Spectrum normalized to $m_E$ against the coupling $\lambda$
as obtained from DLCQ 
for antiperiodic boundary conditions and a harmonic
resolution of K=20. 
{\bf b)} The same spectrum but over a smaller mass range with
larger resolution to emphasise the transition from the discrete
to continuum spectrum.}
\label{spec}
\end{figure}

There are a number of basic features common to all values
of coupling constant. 
The lowest bound state is essentially a fermion-antifermion pair which
becomes the `Schwinger boson' of mass $g/\sqrt{\pi}$
as fermion mass $m_F$ vanishes \cite{Sch62}. For non-zero
fermion mass, this mesonic state has a `size' and so we refer to it
as an `extended' Schwinger boson. Above this state lie several 
bound `molecule' states of the extended Schwinger boson state.
The number of these depend on the strength of the coupling and the
value of the $\theta$ parameter which is essentially a background
electric field \cite{CJS75}. Finally there is a  
a continuous spectrum consisting of extended Schwinger boson
states in relative motion. Embedded within this continuum are
discete bound states of larger `molecules' of extended Schwinger
bosons, for example one of three bosons, namely predominately
a six-particle state \cite{HOT95}. 

\begin{figure}[tbp]
\begin{center}

\begin{picture}(165,60)

\put(0  ,0)   {\bf a)}
\put(55 ,0)   {\bf b)}
\put(110,0)   {\bf c)}



\put(-5,0){\unitlength1mm
 \begin{picture}(50,50)
  \centering
  \makebox[50mm]{
   \epsfxsize=50mm
    \epsffile{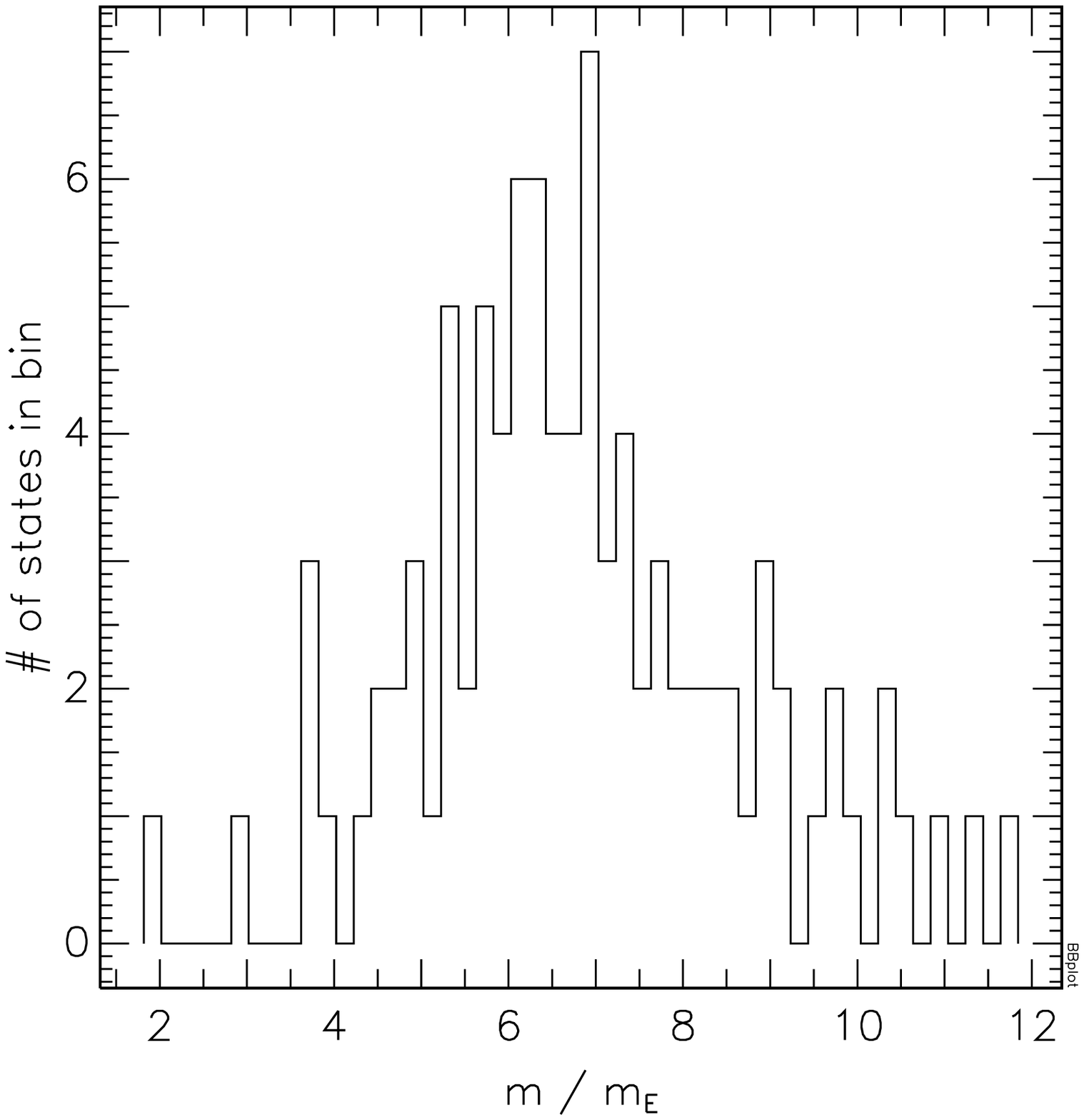}
  }
 \end{picture}
}
\put(50,0){\unitlength1mm
 \begin{picture}(50,50)
  \centering
  \makebox[50mm]{
   \epsfxsize=50mm
    \epsffile{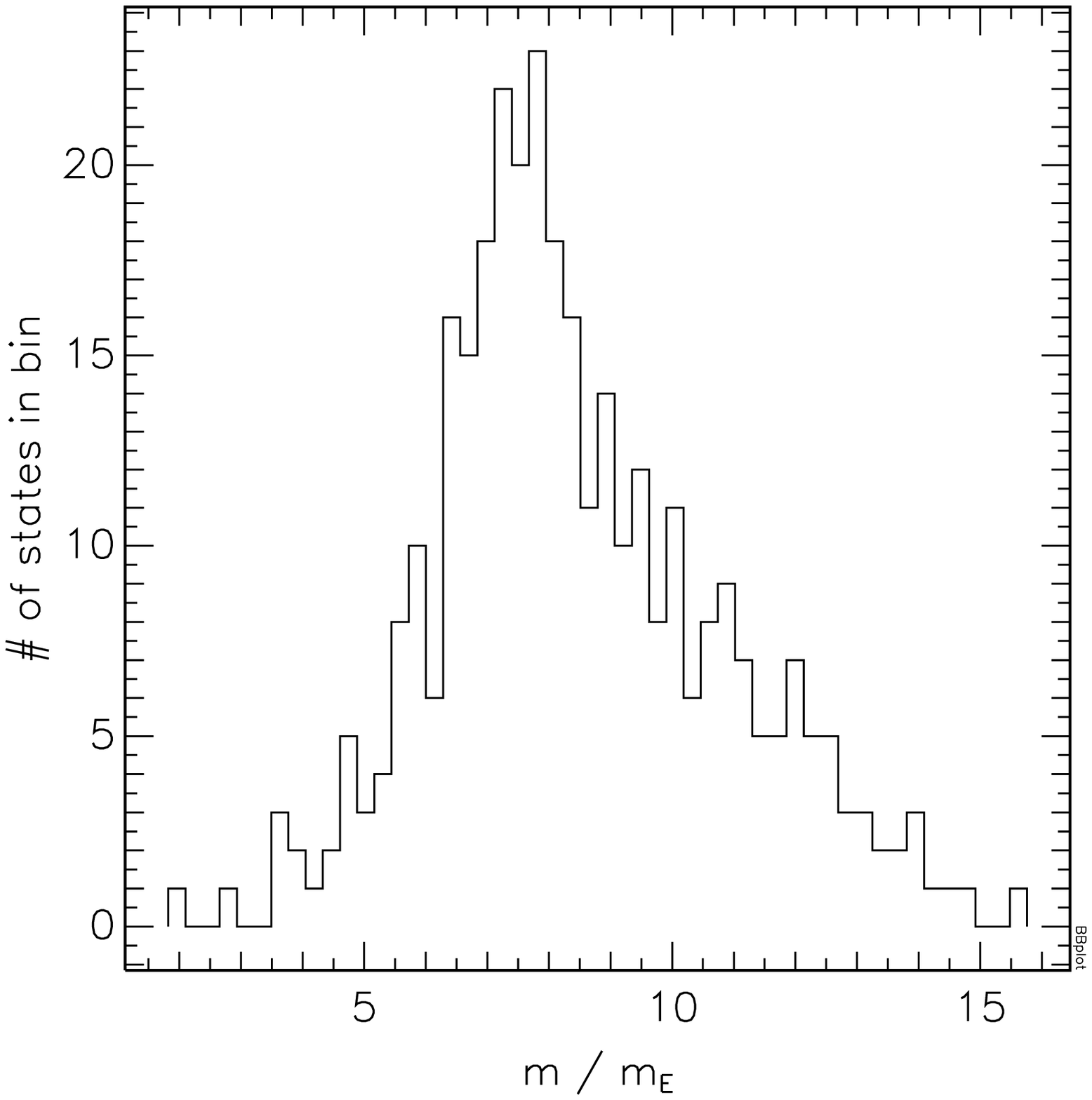}
  }
 \end{picture}
}
\put(105,0){\unitlength1mm
 \begin{picture}(50,50)
  \centering
  \makebox[50mm]{
   \epsfxsize=50mm
    \epsffile{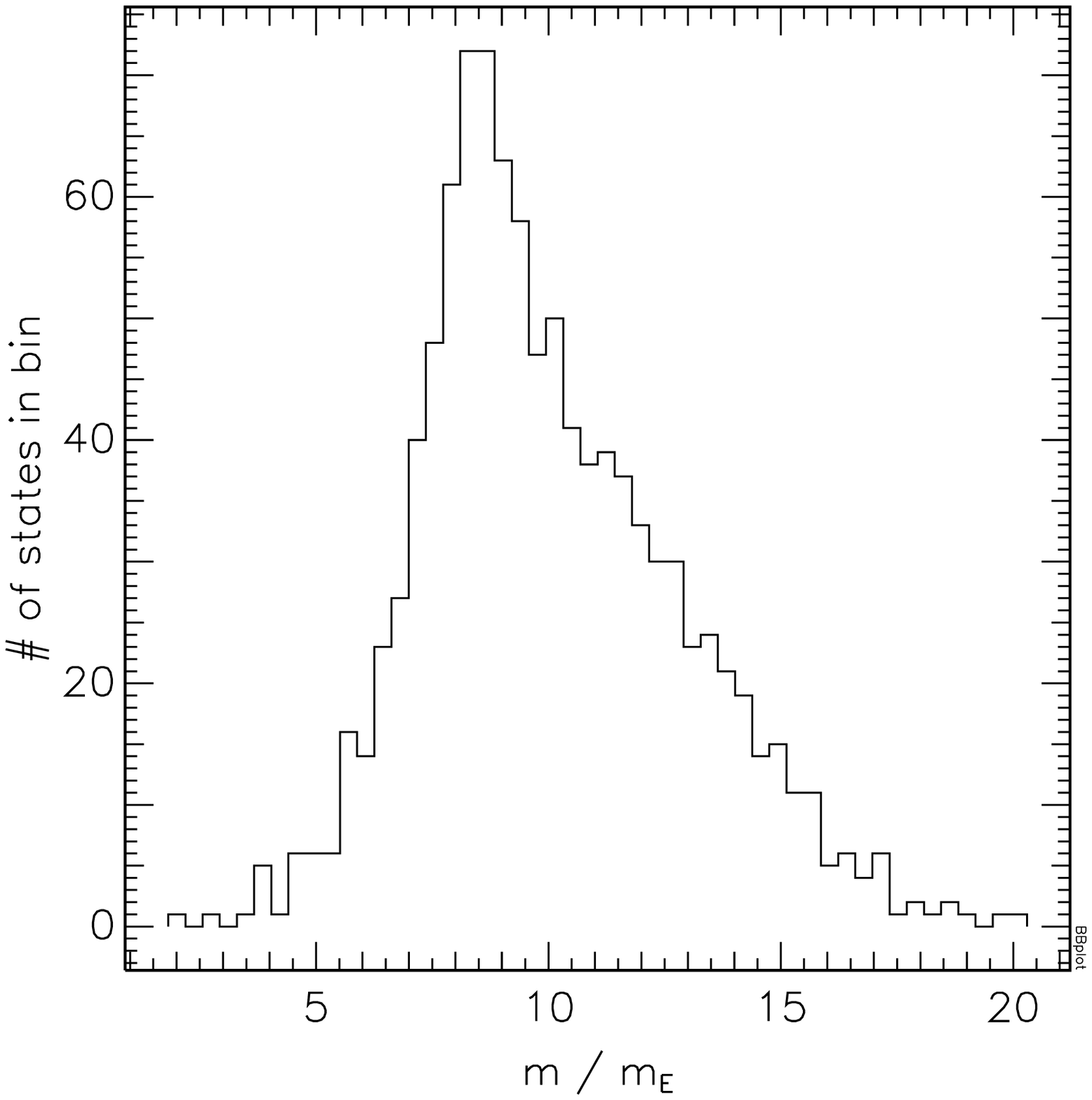}
  }
 \end{picture}
}
\end{picture}
\end{center}

\caption[Density of States]
{{\bf Density of States in Spectrum from DLCQ.}
\sl
The Density of States in the mass spectrum as obtained
from DLCQ with increasing values of harmonic
resolution: (a) K=15, (b) K=20, (c) K=25. The coupling
is relatively strong $\lambda = 0.85$. }
\label{pic0}
\end{figure}

Numerically, for any finite $K$ we always obtain a discrete 
spectrum which in the limit should become a continuum. 
Put another way, for any $K \neq\infty$ there are no scattering states, 
rather states of progressively weaker binding as $K$ increases. 
In particular, the spectrum on its own is rather misleading 
in illustrating how close one is to a continuous spectrum. We give 
the density of states for a particular value of coupling
for various values of $K$ in Fig.\ref{pic0}.

For other couplings the results
are qualitatively similar. One observes that at some point
in the spectrum the density of states begins to decrease
and this is purely an artifact. The point at which the density
begins decreasing does not appear to vary strongly with $K$
within the range of values explored here. It is difficult to
draw any conclusions on this aspect in the $K=\infty$ case.
We do learn here that we should not trust the thermodynamics
for temperatures commensurate with energies
above the point where the density of states decreases.
This will be reflected in the results to be given in the
next section, where we take a rather conservative range of
temperatures.

These features aside, we conclude this section by observing the
basic structure of the density of states: it is basically
a rapidly rising distribution breaking down
at a certain temperature, thus 
roughly resembling a Gaussian with height and width which increase
with increasing $K$. This will be useful at the end in
order to understand certain properties of the thermodynamics
to which we turn now.
    
\section{Canonical Ensemble}
As mentioned, the DLCQ method involves a calculation of
the spectrum for a fixed `size' of system, which is
subsequently varied in order to approach the continuum. 
One should note that this size is not related to that measured in any
conventional frame accessible by proper Lorentz transformations,
and so it is not intuitive to picture this as being in a certain
frame of reference. 
Rather, the method tends to a
system of infinite length in a standard reference frame.
Put another way, the results of DLCQ for, say, a spectrum
acquire relevance to the `real world' with no boundary walls
only when there is some converging
behaviour as $K\rightarrow\infty$. 

Now, as we attempt to describe the thermodynamics of such
a system we are lead to a {\it canonical}
ensemble of such finite size systems, there being
no external source of particles in this problem. Again, these canonical
ensembles bear no resemblance to any finite size system
in a meaningful rest frame. Rather, by taking $K\rightarrow\infty$
we (hope to) approach the {\it same} thermodynamics of a 
set of canonical ensembles of progressively larger volumes
as determined in the usual sense. Thus, it is wrong  
to interpret the results for any {\it finite} $K$. Rather
we must always (where possible) extrapolate at least
qualitatively to the continuum.

We now compute the partition function in the canonical
ensemble using the basic relation 
$Z = \sum_i \exp (- \beta E_i)$ where $E_i$ is the energy measured
in some well-defined rest-frame. We consider the system 
in some given frame attainable by proper Lorentz transformations,  
and now relate the energy $E_i$ to the
relativistic-invariant mass squared of the particles in the
system. 
Here we make our first genuine approximation: we
shall treat all the states over which the partition is sum is
taken as {\it free} discrete bound states,
enabling us to take $M^2$ independent of $P^+$ and to use the 
continuum energy-momentum relation:

\begin{eqnarray}
Z = \sum_{{\rm states}\ i} \  
\sum_{{\rm momenta}\ p_i} e^{- \beta \sqrt{M_i^2 + p_i^2}}
\;.
\label{eq:partition} 
\end{eqnarray}

The momenta $p_i$ are discrete given that the system is taken to be
of finite size $d \neq L$. They are the momenta
of bound states measured in the given
frame for which the size is $d$. 
Eq.(\ref{eq:partition}) is valid for the low energy
spectrum which is indeed that of free bound states, either
of a single extended Schwinger boson, or the molecule 
states. We now use for $M_i^2$ the values obtained
from a DLCQ calculation for a given coupling $\lambda$
and harmonic resolution $K$. Insofar as for any finite $K$ all the 
DLCQ states {\it are} discrete, Eq.(\ref{eq:partition}) 
would appear to be correct.  
Where the error lies is in concluding from finite $K$ results  
the thermodynamic properties of the finite size system in
some Lorentz frame. Unfortunately, at this point we
cannot determine the error being made following this route.

From Eq.(\ref{eq:partition}) the procedure is now straightforward.    
A useful trick is to approximate the momentum sum by an
integral,   
\begin{eqnarray}
Z &=& {d \over 2\pi} \sum_i 2 
\int_0^{\infty}  \,dp_i e^{- \beta \sqrt{M_i^2 + p_i^2}} \nonumber \\
&=&{d \over \pi} \sum_i M_i K_1 (\beta M_i) 
\end{eqnarray}
with $K_1$ the first Bessel function in the notation of \cite{Grad65},
where we add that
integrating over $P^+$ in the same approximation
results of course in the same expression. 
We take Boltzmann's constant to be one so that $T = \beta^{-1} $. 
The energy $ e $ and specific heat $ c $
are calculated via the relations
\begin{eqnarray}
e  &=& {\partial ln Z \over \partial T} \cdot T^2 \quad {\rm and} 
\nonumber \\
c  &=& {\partial e \over \partial T}.
\end{eqnarray}

\begin{figure}[tbp]

\begin{picture}(150,70)
\put(0  ,0)   {\bf a)}
\put(55 ,0)   {\bf b)}
\put(110,0)   {\bf c)}



\put(-5,0){\unitlength1mm
 \begin{picture}(50,50)
  \centering
  \makebox[50mm]{
   \epsfxsize=50mm
    \epsffile{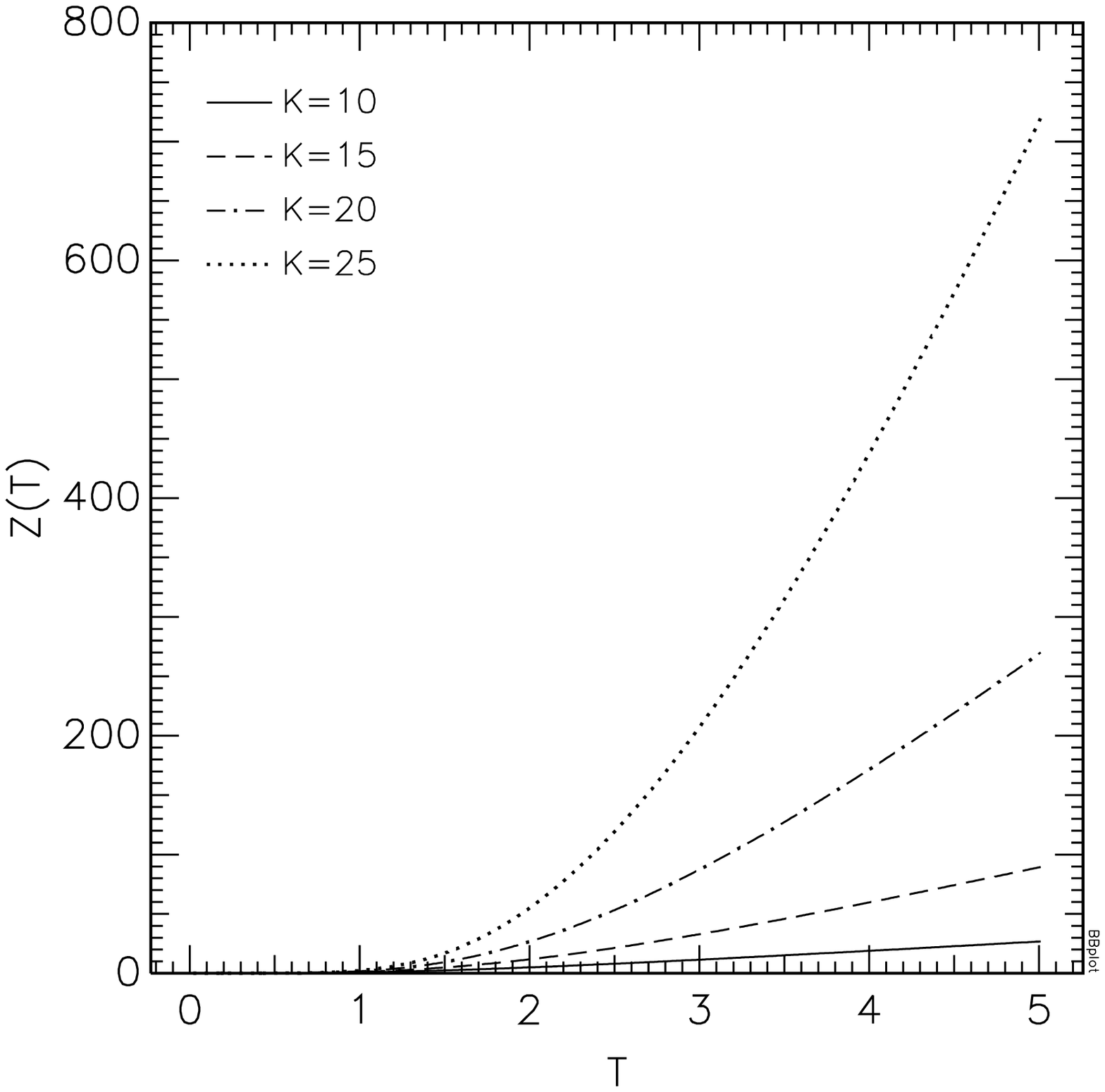}
  }
 \end{picture}
}
\put(50,0){\unitlength1mm
 \begin{picture}(50,50)
  \centering
  \makebox[50mm]{
   \epsfxsize=50mm
    \epsffile{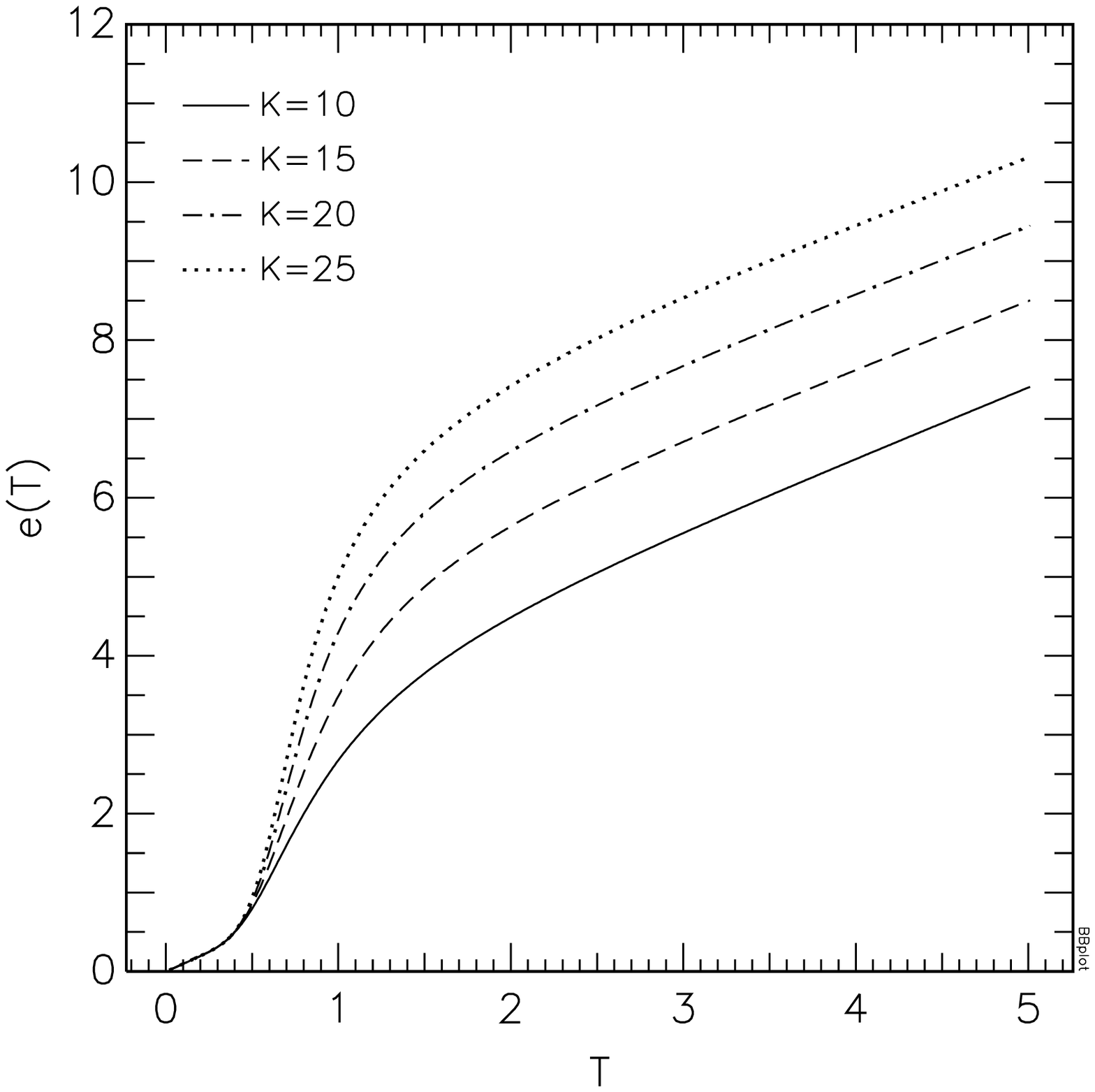}
  }
 \end{picture}
}
\put(105,0){\unitlength1mm
 \begin{picture}(50,50)
  \centering
  \makebox[50mm]{
   \epsfxsize=50mm
    \epsffile{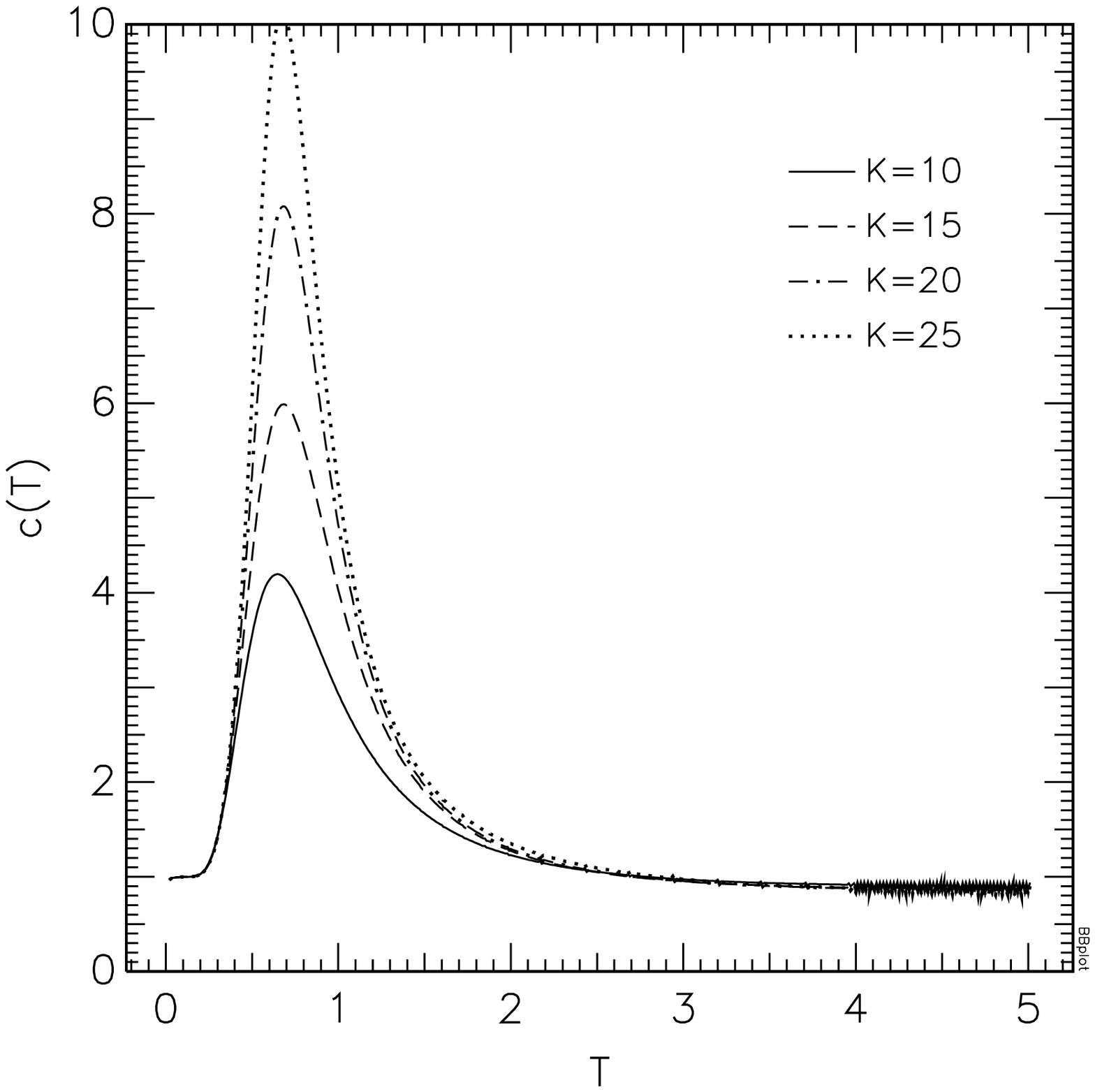}
  }
 \end{picture}
}
\end{picture}

\caption[Partition function, Energy and Specific Heat]
{{\bf Partition function, Energy and Specific Heat.}
\sl  
Results of a DLCQ calculation with antiperiodic 
boundary conditions and coupling $\lambda = 0.30$
for the following quantities: 
{\bf a)} Partition function,
{\bf b)} Energy,
{\bf c)} Specific heat. 
}
\label{pic1}
\end{figure}

Now, given the spectrum for a given $K$ and $\lambda$
the above quantities are straightforwardly calculated.
We first present the results for an intermediate
value of coupling and various values of $K$ in Fig.\ref{pic1}.

The most significant feature we observe is the rising peak in
the specific heat as $K$ increases. This region aside,
for other temperatures the different $K$ values coincide.
Secondly, we see a rise in the energy above this peak temperature
with increasing K.

\begin{figure}[tbp]
\begin{picture}(160,60)
\put(0  ,0)   {\bf a)}
\put(55 ,0)   {\bf b)}
\put(110,0)   {\bf c)}


\put(-5,0){\unitlength1mm
 \begin{picture}(50,50)
  \centering
  \makebox[50mm]{
   \epsfxsize=50mm
    \epsffile{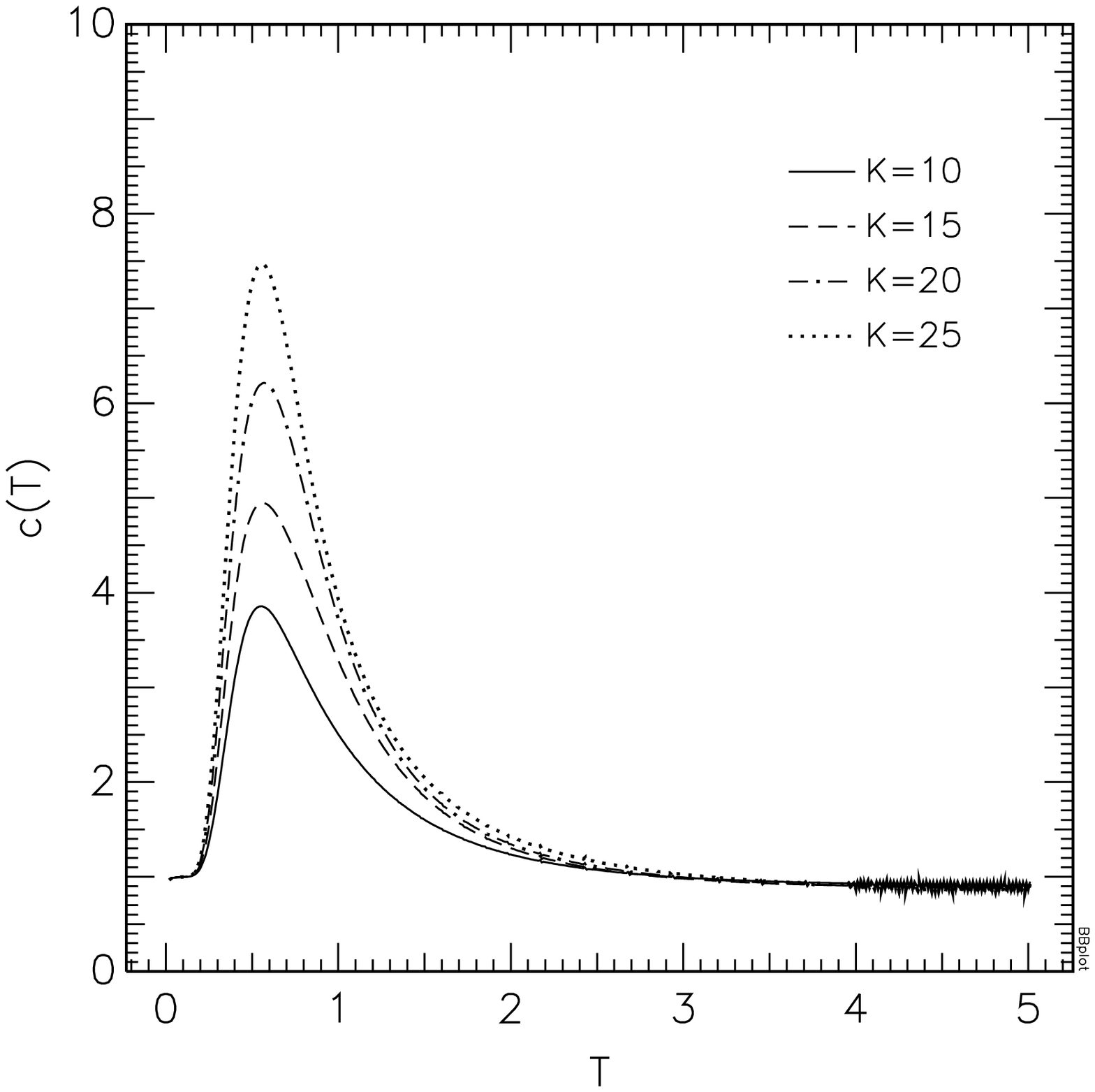}
  }
 \end{picture}
}
\put(50,0){\unitlength1mm
 \begin{picture}(50,50)
  \centering
  \makebox[50mm]{
   \epsfxsize=50mm
    \epsffile{specific-0.30.ps}
  }
 \end{picture}
}
\put(105,0){\unitlength1mm
 \begin{picture}(50,50)
  \centering
  \makebox[50mm]{
   \epsfxsize=50mm
    \epsffile{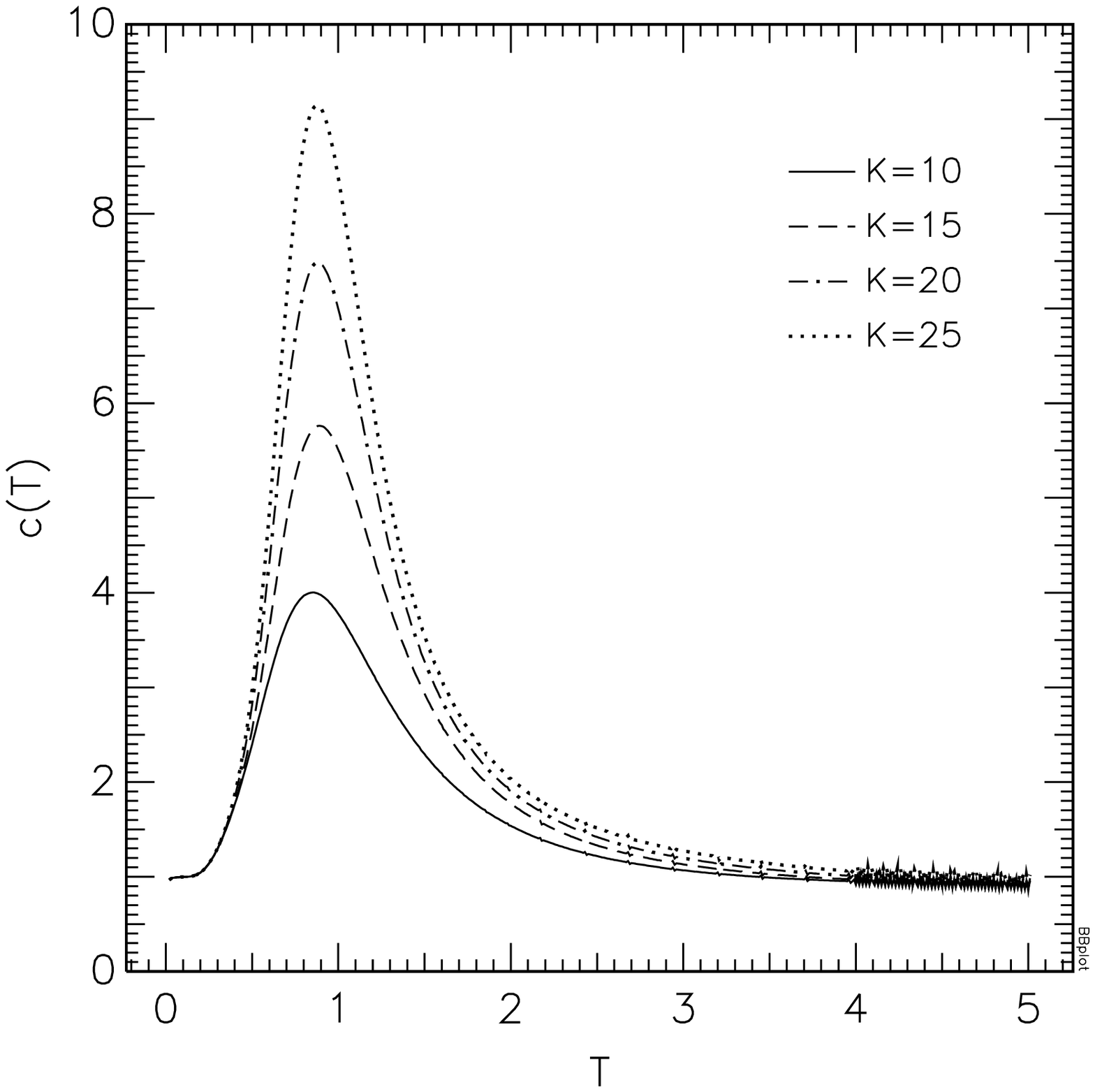}
  }
 \end{picture}
}
\end{picture}

\caption[Specific Heats for Different Couplings]
{{\bf Specific Heat for Three Values of Coupling.}
\sl
The specific heat for couplings {\bf a)} $\lambda=0.05$,
{\bf b)} $\lambda = 0.30$, {\bf c)} $\lambda = 0.85$. 
}
\label{pic2}
\end{figure}

For different couplings, the results qualitatively do not
change. Quantitatively, the changes are most evident
by looking just at the specific heat. In Fig.\ref{pic2}
we compare this for three different couplings. 
The only slight change is in the position and height of
the peak as the coupling increases. More precisely,
inspection shows the temperature at which the peak
occurs appears to approach $T=1$ as $\lambda \rightarrow 1$,
which coincides with the strong coupling or alternately
massless fermion case. We mention that in our units
the lowest state in this regime, the Schwinger boson,
has mass $M=1$.   

\section{Discussion}
Before attempting to understand these results, we first
note that we have given values in temperature which do not
correspond to systems in the ensemble with states
occupying deeply into the putative continuum spectrum.
Insofar as, with increasing $K$, we trust the low energy
derived spectrum the only error we could be making is in
treating even the low part of the scattering spectrum
on the same footing as the discrete one. As mentioned,
we are unable at this point to quantify how significant this error is. 

Can we nonetheless understand the origin of the peak? 
Based on our earlier observation that the density of
states roughly resembles a Gaussian distribution, 
we can try simple analytic calculations using 
\begin{equation}
\rho(E) = \exp(-(E-E_0)^2/A)
\; .
\end{equation}
This indeed gives a qualitatively similarly increasing peak in
the specific heat with {\it decreasing} $A$.
In other words, the height of the peak comes from
the steepness of the rise in the density of states
as $K$ increases.

The question now is: can these results be extrapolated to the
continuum to draw physically relevant information? 
This is impeded by several features.
The first is a conceptual problem: relating $K$ to a physical length
scale for a finite size scaling analysis. 
The second impediment is entirely a practical one:  
going beyond the present values of $K$ while
maintaining quick CPU time. The question is 
whether there is some convergence to
a finite value for the specific heat at the turning point, or
whether indeed it diverges to infinity.
The associated question is whether the profile of the rise in
the density of states actually converges to some fixed form
in the continuum limit.
Either way, increased computing power
is necessary which could be complemented by algorithms such
as the Lanczos method \cite{Lan50}. Only via such computations
could we say with any confidence whether the peak is
indicative of a phase transition. 

Let us, nonetheless, {\it assume} scaling behaviour,
and use the values of the peak in the specific
heat to attempt a numerical extrapolation over increasing 
values of $K$. In this way we have estimated the critical exponent $\alpha$ 
(see, for example, \cite{Bin92}) associated
with the suggested second-order phase transition. 
Direct fits ranging from $K=10$ to $25$ resulted in
increasing exponents from $\alpha \approx 0.38$ to $0.61$.    
At best, if this
is critical behaviour we  
estimate a lower bound for the exponent 
at $\alpha > 0.7$.
A scaling analysis using $K$ as our relevant length scale
(in the absence of any other choice) 
resulted in a determination of $\alpha$ and $T_c$
for $\lambda=0.05$ to $\alpha = 0.89 \pm 0.04$
and $T_c=0.54 \pm 0.04$
consistent with the above bounds.
Of course, all this is severely limited by
the problems discussed above but we put these
estimates forward as the basis for further discussion. 


At any rate, for low to moderate temperatures the method of
calculation would appear to be consistent and yields 
physically sensible results. An interesting  
calculation which could be pursued with the present method
is that of the temperature
dependence of the chiral condensate.
There exist analytic results in the literature which
allow some comparison \cite{Smi92}.

In summary, we have presented a method for extracting
thermodynamic quantities from the spectrum of a given
field theory as computed using Discretised Light-Cone
Quantisation. The particular example we chose was
QED in (1+1) dimensions. Were the DLCQ program for QCD to
be achieved, namely the generation of a hadron spectrum,
then the method presented here could be generalized
with little effort. 
Of course, computing the spectrum
of QCD from a `first principles' calculation remains
the difficult challenging task, but nonetheless
the extension of the method proposed in the
present work will be essential for such computations.

\section*{Acknowledgements}
We are grateful to L.C.L. Hollenberg for the original suggestion
and urging to pursue this project. Also, J. Vary, D.G. Robertson,
B. Bunk,
H.-J. Pirner, H.-C. Pauli, S.P. Klevansky, J. Rau, P-B. Gossiaux 
and J. H\"ufner are all thanked for 
profitable discussions.
SE thanks K. Riechmann
for technical help during the preparation and the 
Max-Planck-Institut f\"ur Kernphysik for its hospitality at stages
of preparation of this work.
ACK was supported by a Max-Planck-Gesellschaft Stipendium,
SE by Deutsche Forschungsgemeinschaft grant No. 
WO 389/3-2.
 

\end{document}